\documentclass[12pt,reqno]{amsproc}

\usepackage{amsmath,amsfonts,amsthm,amssymb}

\usepackage{amsaddr}

\usepackage[a4paper,height=20cm,top=24mm,bottom=24mm,left=26mm,right=26mm]{geometry}

\usepackage[numbers,sort]{natbib}
\usepackage[breaklinks=true]{hyperref}

\usepackage[footnotesize,hang,sf]{caption}\setcaptionwidth{12cm}

\usepackage[]{graphicx}            
\usepackage{booktabs}
\usepackage[table]{xcolor}

\DeclareMathAlphabet{\mathpzc}{OT1}{pzc}{m}{it}
\usepackage{titlesec}
\titleformat{\section}
{\bfseries\scshape\centering}
{\thesection.}{.5em}{}
\titleformat{\subsection}
{\rmfamily\bfseries}
{\thesubsection}{.5em}{}
\titleformat{\subsubsection}
{\rmfamily\bfseries}
{\thesubsubsection}{.5em}{}
\numberwithin{equation}{section}

\newcommand{\I}{\mathrm{i}}

\DeclareMathOperator{\ch}{ch}

\DeclareMathDelimiter{\Norm}{\mathord}{largesymbols}{"3E}{largesymbols}{"3E}

\allowdisplaybreaks[4]
\begin{document}
\baselineskip 16pt
\parskip 8pt
\sloppy


\title{Enriques Moonshine}


\author[T. Eguchi]{Tohru \textsc{Eguchi}}

\address{Department of Physics and Research Center for Mathematical
  Physics,
  \\
  Rikkyo University,
  Tokyo 171-8501, Japan.}
\email{\texttt{tohru.eguchi@gmail.com}}


\author[K. Hikami]{Kazuhiro \textsc{Hikami}}
\address{Faculty of Mathematics,
  Kyushu University,
  Fukuoka 819-0395, Japan.}

\email{\texttt{khikami@gmail.com}}


\date{May 6, 2013.}

\begin{abstract}
We propose a new moonshine phenomenon associated with the elliptic
genus of the Enriques surface 
($1\over 2$ of the elliptic genus of $K3$)
with the symmetry group given by the Mathieu group $M_{12}$.
\end{abstract}





\maketitle


\section{\mathversion{bold}Mathieu moonshine}

Recently a new moonshine phenomenon associated with the elliptic genus
of $K3$ surface  has been discovered and is receiving some
attentions.
It was first observed in~\cite{EgucOoguTach10a}
that when one expands the elliptic genus of $K3$ in terms of
irreducible characters of $\mathcal{N}=4$ superconformal algebra
(SCA) the expansion coefficients $A(n)$ at lower values of $n$ are
decomposed into a sum of dimensions of irreducible representations
(irreps.) of the Mathieu group $M_{24}$.
Subsequently the twisted elliptic genera of $K3$ surface for each conjugacy class $g$ of
$M_{24}$
(analogues of McKay--Thompson series of monstrous moonshine)
have been constructed and used to determine systematically the
decomposition of expansion coefficients up to very high values of $n$
($\sim 1000$)~\cite{MCheng10a,  GabeHoheVolp10a,GabeHoheVolp10b,EguchiHikami10b}.
Finally a mathematical proof has been given to show that expansion
coefficients are in fact decomposed into a sum of dimensions of
irreps. of $M_{24}$ with positive and integral multiplicities for all
values of~$n$~\cite{Gannon12a}.  
Thus the ``Mathieu moonshine'' phenomenon has now been established
although its physical or mathematical origin is not yet explained.

We present the character table and list of conjugacy classes of $M_{24}$ 
in Tables~\ref{tab:character_M24} and~\ref{tab:permutation_M24}.
We also present the data of the decomposition of expansion coefficients
$A(n)$  of elliptic genus of $K3$ 
\begin{equation}
  Z^{K3}(z;\tau)
  =24 \ch^{\widetilde{R}}_{h={1\over 4},{\ell}=0}(z;\tau)
  +\sum_{n=0}^{\infty}A(n) \ch^{\widetilde{R}}_{h=n+{1\over
      4},\ell={1\over 2}}(z;\tau)
\end{equation}
into irreps. of $M_{24}$
in Table~\ref{tab:multiplicity_M24}.
Note that here $Z^{K3}$ denotes the elliptic genus of $K3$ and 
$\ch_{h={1\over 4}, \ell}^{\widetilde{R}}$ and 
$\ch_{h=n+{1\over   4}, \ell}^{\widetilde{R}}$ are massless (BPS) and massive (non-BPS)
characters (with $h=n+{1\over 4}$ and spin-$\ell$) of ${\mathcal N}=4$ SCA in
 R-sector with $(-1)^F$ insertion. For later use we also record the
 data of expansion coefficients $A_g(n)$ of twisted elliptic genera
 $Z^{K3}_g(z;\tau)$ of $K3$ for each conjugacy class $g \in M_{24}$
 \begin{equation}
 Z^{K3}_g(z;\tau)
 =\chi_g \ch^{\widetilde{R}}_{h={1\over     4},{\ell}=0}(z;\tau)
 +
 \sum_{n=0}^{\infty}A_g(n)
 \ch^{\widetilde{R}}_{h=n+{1\over  4},\ell={1\over 2}}(z;\tau) ,
\end{equation}
in Table~\ref{tab:coeff_M24}.
Note that $A(n)\equiv A_{\mathrm{1A}}(n)$.
 
Recently there has been an attempt at generalizing Mathieu moonshine~\cite{ChenDuncHarv12a}
based on suitable Jacobi forms with higher values of indices $>1$ and
again expanding them in terms of $\mathcal{N}=4$ superconformal
characters using the
data of~\cite{EguchiHikami09b}. 
This ``umbral moonshine'' sequence has smaller symmetry groups than
$M_{24}$.
Unfortunately, its Jacobi forms do not correspond to elliptic genera of  
any complex manifolds and the connection to geometry is not clear in
umbral moonshine.
In~\cite{EguchiHikami12a} we have discussed a still another example of
moonshine based on $\mathcal{N}=2$ SCA instead of
$\mathcal{N}=4$.

\section{Enriques moonshine}

In this paper we want to propose a new example of moonshine
phenomenon which may be called as ``Enriques moonshine''.
It is defined  by the elliptic genus of Enriques surface expanded in
terms of $\mathcal{N}=4$ characters. Its 
symmetry group is $M_{12}$.
Recall that Enriques surface is closely related to $K3$:
it is  obtained by quotienting $K3$ by a fix-point free involution and
has an Euler number $12$.
Its elliptic genus is one half of that of $K3$
\begin{equation}
  Z^{\text{Enriques}}(z;\tau)
  =
  {1\over 2}Z^{K3}(z;\tau)
  =
  4\left[
    \left({\theta_{10}(z;\tau)\over \theta_{10}(0;\tau)}\right)^2
    +
    \left({\theta_{00}(z;\tau)\over \theta_{00}(0;\tau)}\right)^2
    +
    \left({\theta_{01}(z;\tau)\over \theta_{01}(0;\tau)}\right)^2
  \right] .
\end{equation}

Enriques moonshine is motivated by the following simple
considerations.
\begin{enumerate}
\def\labelenumi{\theenumi.}
\item 
  It is known that in the case of Mathieu moonshine the expansion
  coefficients $A(n)$    are always even for any $n\ge 1$:
  this is because (i) when the decomposition of $A(n)$ contains a
  complex representation of $M_{24}$, 
  it also contains its complex conjugate representation,
  and
  (ii) when $A(n)$ contains a real representation its multiplicity is
  always even~\cite{Gannon12a}.
  
\item
  Thus in order to keep integrality of the decomposition when  we
  divide by $2$ the $K3$ elliptic genus we just need to find a subgroup $G$
  of $M_{24}$ where all the complex representations of $M_{24}$ become
  real representations of $G$.
  It turns out that this is the case of $M_{12}$.
  
\item
  Geometrical considerations on Enriques surface suggests the
  relevance of the symmetry group $M_{12}$~\cite{Mukai12a}.

\end{enumerate}

Let us first derive the decomposition of $M_{24}$ representations (reps.) into
those of  $M_{12}$ in order to examine the reality of representations.
For this purpose we want to make a correspondence between the
conjugacy classes of  the two groups.
In Table~\ref{tab:permutation_M12} we list the conjugacy classes of
$M_{12}$ and their permutation representations.
We recall that Mathieu group $M_{24}$ is the symmetry group of Golay
code and permutes dodecads into each other.
$M_{12}$ is the subgroup
of $M_{24}$  which fixes a dodecad~\cite{Conway98a}. 
Conjugacy class of $\mathrm{2A}$ of $M_{12}$, for instance, has a cycle shape
$2^6$ and it is natural that this  corresponds to the conjugacy class
$\mathrm{2B}$ of $M_{24}$ with a cycle shape $2^{12}$.
Thus in general a class $g$ of $M_{12}$ 
should correspond to a class $g^\prime$ of $M_{24}$ whose cycle shape is the
square of that of $g$.
There are exceptions to this rule when there
exists a non-trivial outer automorphism between conjugacy classes of $M_{12}$.
From the Table~\ref{tab:permutation_M12} we note that the sizes of
conjugacy classes 
are equal for the pair $\mathrm{4A}, \mathrm{4B}$ and
$\mathrm{8A}, \mathrm{8B}$ and
$\mathrm{11A}, \mathrm{11B}$.  
It is known~\cite{Conway98a} that  these pairs are tied by a non-trivial outer
automorphism~$\sigma$.
If one takes a class $g$ of $M_{12}$ 
the corresponding class of $M_{24}$ should become $g\cup\sigma(g)$.  
In the case of $g=\mathrm{4A}$,
$\sigma(\mathrm{4A})=\mathrm{4B}$, for instance,
the cycle shape of $g \cup\sigma(g)$ equals $4^22^2 \cup 4^21^4$
and that of $g^\prime$ becomes $4^42^21^4$ which is class
$\mathrm{4B}$ of $M_{24}$.
Thus $\mathrm{4A}, \mathrm{4B}$ of $M_{12}$ both should correspond to
$\mathrm{4B}$ of $M_{24}$.
In this way we can construct the following table of correspondences.  
\begin{equation}
  \label{g_and_prime}
  \newcolumntype{L}{>{$}l<{$}}
  \newcolumntype{R}{>{$}r<{$}}
  \newcolumntype{C}{>{$}c<{$}}
  \centering
  \begin{tabular}[]{L*{16}{C}}
    \toprule
    g  \in M_{12}&
    \mathrm{1A} &\mathrm{2A} &\mathrm{2B} &\mathrm{3A} &\mathrm{3B} &
    \mathrm{4A} &\mathrm{4B} &\mathrm{5A} &\mathrm{6A} &\mathrm{6B} &
    \mathrm{8A} &\mathrm{8B} &\mathrm{10A} &\mathrm{11A} &\mathrm{11B} 
    \\
    \midrule
    g^\prime \in M_{24}    &
    \mathrm{1A} &\mathrm{2B} &\mathrm{2A} &\mathrm{3A} &\mathrm{3B} &
    \mathrm{4B} &\mathrm{4B} &\mathrm{5A} &\mathrm{6B} &\mathrm{6A} &
    \mathrm{8A} &\mathrm{8A} &\mathrm{10A} &\mathrm{11A} &\mathrm{11A} 
    \\
    \bottomrule
  \end{tabular}
\end{equation}

Let us now determine the branching rule of the irreps. of $M_{24}$  into those of $M_{12}$.
We consider the following ``inner product'' of character tables of $M_{24}$ and $M_{12}$
to derive the multiplicity of a representation $r$ of $M_{12}$ contained 
in the  
representation~$R$ of $M_{24}$
\begin{equation}
  \sum_{g}
  \chi(M_{24})_{R}^{\,\,\,\,  t(g)}
  {\chi(M_{12})^{~~-1}}_{g}^{\,\,\,r}
  =
  \text{multiplicity of irrep. $r$   in irrep. $R$}
\end{equation}
Here $t(g)=g^\prime$ of~\eqref{g_and_prime}, and
$\chi(M_{12})^{-1}$ is the inverse of the character
table of $M_{12}$ in the sense of a matrix.
Using the character tables of $M_{24}$, $M_{12}$ in
Tables~\ref{tab:character_M24},~\ref{tab:character_M12},
we find the above 
multiplicities as given by Table~\ref{tab:branching}. 
Note that as we mentioned already, decomposition of complex
representations of $M_{24}$ contain only real representations of
$M_{12}$ or the sum of
pairs of
complex conjugate representations of  $M_{12}$.

Therefore if we substitute $M_{24}$ reps. by their 
$M_{12}$ decompositions in the Mathieu
moonshine of Table \ref{tab:multiplicity_M24}, and divide by an
overall factor 2, we maintain the integrality of multiplicities of
$M_{12}$ representations.
One obtains the decomposition of the elliptic genus of  
Enriques surface  given in terms of  $M_{12}$ reps.
See Table~\ref{tab:multiplicity_M12}.

There is in fact a more elegant way to derive the decomposition of Enriques elliptic genus.
This is to use the method of twisted elliptic genus.
We have at hand the twisted genera for all conjugacy classes in
Mathieu moonshine (tabulated in \cite{EguchiHikami10b}) and we can use these results.
We introduce an ansatz that twisted elliptic genera for Enriques moonshine are one half
of those of Mathieu moonshine of the corresponding conjugacy classes 
\begin{align}
  Z_{g}^{\text{Enriques}}(z;\tau)
  & =
  \frac{1}{2} Z_{t(g)}^{K3} (z;\tau)
  \hskip1cm \text{for all conjugacy classes $g$ of $M_{12}$}
\end{align}
 
Then by introducing the expansion coefficients $A^{\text{Enriques}}_g(n)$ for all 
classes $g \in M_{12}$
\begin{align}
  Z_g^{\text{Enriques}}(z;\tau)
  =
  \chi_g^{\text{Enriques}} \ch_{h={1\over 4},\ell=0}(z;\tau)
  +
  \sum_{n=0}^{\infty}A_g^{\text{Enriques}}(n) \ch_{h=n+{1\over 4},\ell={1\over 2}}(z;\tau)
\end{align}
where
$\chi_g^{\text{Enriques}}$ is the Euler number
$\chi_g^{\text{Enriques}}=Z_g^{\text{Enriques}}(0;\tau)$,
\begin{equation*}
  \newcolumntype{L}{>{$}l<{$}}
  \newcolumntype{R}{>{$}r<{$}}
  \newcolumntype{C}{>{$}c<{$}}
  \centering
  \begin{tabular}[]{L*{16}{C}}
    \toprule
    g  \in M_{12}&
    \mathrm{1A} &\mathrm{2A} &\mathrm{2B} &\mathrm{3A} &\mathrm{3B} &
    \mathrm{4A} &\mathrm{4B} &\mathrm{5A} &\mathrm{6A} &\mathrm{6B} &
    \mathrm{8A} &\mathrm{8B} &\mathrm{10A} &\mathrm{11A} &\mathrm{11B} 
    \\
    \midrule
    \chi_g^{\text{Enriques}} &
    12 & 0 & 4 & 3 & 0 & 2 & 2 & 2 & 0 & 1 & 1 & 1 & 0 & 1 & 1
    \\
    \bottomrule
  \end{tabular}
\end{equation*}
we obtain the multiplicity for the $M_{12}$ representation $r$ at level $n$  
\begin{equation}
  \sum_g {n_g \over |G|}
   \,   {\overline{\chi(M_{12})}}_r^{\,\,\,g} \, A^{\text{Enriques}}_g(n)
  =c^{\text{Enriques}}_r(n).
\end{equation}
Here $|G|$ denotes the order of $M_{12}$(=95040) and $n_g$ is the size of $M_{12}$ conjugacy class $g$.\\

By using the orthogonality relation of the character table it is possible to prove that the above formula in fact reproduces the data of Table~\ref{tab:multiplicity_M12}. First we recall that the multiplicity of representation $R$ in Mathieu moonshine is given by 
\begin{eqnarray}
&&\sum_{g'} {n_{g'} \over |G^\prime|} \, 
\overline{\chi(M_{24})}_R^{\,\,\,g'} \, A_{g'}(n)=c_R^{K3}(n)
\end{eqnarray}
Here $g'$ runs over conjugacy classes of $M_{24}$,
and $|G^\prime|$ is the  order of $M_{24}$.
We covert $M_{24}$ representations into $M_{12}$ representations and divide by 2 to obtain multiplicities in Enriques moonshine 
 \begin{eqnarray}
&&c_r^{\text{Enriques}}(n)
={1\over 2}\sum_R
\left[\sum_{g'} {n_{g'}\over  |G'|} 
  \overline{\chi(M_{24})}_R^{\,\,\,g'} \, A_{g'}(n)\right]\times 
\left[\sum_{g} \chi(M_{24})_R^{\,\,\,t(g)} \,
  (\chi(M_{12})^{-1})_g^{\,\,\,r}\right]\nonumber \\
&&={1\over 2}\sum_{g,g'}\delta(g',t(g)) \,
(\chi(M_{12})^{-1})_g^{\,\,\,r} \, A_{g'}(n)
=
{1\over 2}\sum_g {n_g\over |G|} \, \overline{\chi(M_{12})}_r^{\,\,\,g}
\,
A_{t(g)}(n)\nonumber \\
&&=
\sum_g {n_g\over |G|}\, \overline{\chi(M_{12})}_r^{\,\,\,g} \,
A_g^{\text{Enriques}}(n).
\end{eqnarray}

\section{Discussions}

In this article we have taken one half of the elliptic genus of $K3$ and obtained the 
Enriques moonshine.
Consistency of the Enriques surface as a string theory background is a  
delicate issue since its
canonical class
does not quite vanish while it
carries a Ricci flat K\"ahler metric.
We do not consider such questions 
in this paper and are primarily concerned with the possibility of the action of
symmetry group on the elliptic genus $Z^{\text{Enriques}}$. 

We have shown that $M_{12}$ in fact acts on $Z^{\text{Enriques}}$. 
We should note, however, that a symmetry group still larger than $M_{12}$ may possibly act on the elliptic genus. 
We have evidence that a maximal subgroup $M_{12}\hskip-1mm : \hskip-1mm 2$
of $M_{24}$ (binary extension of $M_{12}$) acts on $Z^{\text{Enriques}}$.  

It was crucial for the existence of Enriques moonshine that all the multiplicities of real representations of $M_{24}$ are even  integers in Mathieu moonshine.
We have recently noticed that similar phenomena take place in the Umbral moonshine and thus it is quite likely 
that we can take one half of Jacobi forms of Umbral moonshine and construct a new moonshine series with reduced symmetry groups. 
This issue will be discussed in a forthcoming publication~\cite{EguchiHikami13}.

\bigskip
\noindent
\textbf{Note added:} After the original version of this paper has been submitted to arXiv we came to
know the paper~\cite{Govind10b} by S.~Govindarajan where the group
$M_{12}$ is used as the symmetry group of Mathieu moonshine.
In this paper the relation~\eqref{g_and_prime} between the conjugacy
classes of $M_{24}$ and $M_{12}$ has been obtained.
Also the multiplicities of irreps. of $M_{12}$ in the decomposition of
expansion coefficients $A(n)$ at smaller values of $n$ have been
obtained in agreement with our results of Enriques moonshine upto an
overall factor $2$.
We thank S.~Govindarajan  for informing us of this paper.

\section*{Acknowledgments}

T.E. would like to thank Y.Tachikawa for discussions on the relation
between $M_{24}$ and $M_{12}$. He also thanks discussions with
S. Mukai on Enriques surface. Research of T.E. is supported in part by JSPS KAKENHI Grant Number 22224001, 23340115.
Research of K.H. is supported in part by JSPS KAKENHI
Grant Number 23340115, 24654041.



\begin{table}[h]
  \newcolumntype{L}{>{$}l<{$}}
  \newcolumntype{R}{>{$}r<{$}}
  \newcolumntype{C}{>{$}c<{$}}
  \rowcolors{2}{gray!13}{}
  \centering
  \rotatebox[]{90}{
    \resizebox{.98\textheight}{!}{
      \begin{tabular}[]{R|*{26}{R}}
        \toprule
        &
        \mathrm{1A}&        
        \mathrm{2A}&
        \mathrm{2B}&  
        \mathrm{3A} & \mathrm{3B}&
        \mathrm{4A} &
        \mathrm{4B}& \mathrm{4C}&
        \mathrm{5A}&                
        \mathrm{6A}& \mathrm{6B}&
        \mathrm{7A}& \mathrm{7B}&
        \mathrm{8A}&
        \mathrm{10A} &
        \mathrm{11A}&
        \mathrm{12A} & \mathrm{12B} &
        \mathrm{14A} & \mathrm{14B} & 
        \mathrm{15A} & \mathrm{15B} &
        \mathrm{21A} & \mathrm{21B} &
        \mathrm{23A} & \mathrm{23B}
        \\
        \midrule\midrule
        \chi_1 &
        1 & 1 & 1 & 1 & 1 & 1 & 1 & 1 & 1 & 1 & 1 & 1 & 1 & 1 & 1 &
        1 & 1 & 1 & 1 & 1 & 1 & 1 & 1 &   1 & 1 & 1
        \\
        \chi_2 &
        23 & 7 & -1 & 5 & -1 & -1 & 3 & -1 & 3 & 1 & -1 & 2 & 2 & 1
        & -1 & 1 & -1 & -1 & 0 &   0 & 0 & 0 & -1 & -1 & 0 & 0
        \\
        \chi_3 &
        45 & -3 & 5 & 0 & 3 & -3 & 1 & 1 & 0 & 0 & -1 &
        \frac{-1 + \I \sqrt{7}}{2} & \frac{-1 - \I \sqrt{7}}{2} & -1 & 0 & 1
        & 0 &  1 &
        -\frac{-1 + \I \sqrt{7}}{2} & -\frac{-1 - \I \sqrt{7}}{2} & 0 & 
        0 & \frac{-1 + \I \sqrt{7}}{2} & \frac{-1 - \I \sqrt{7}}{2} &
        -1 & -1
        \\
        \chi_4 &
        45 & -3 & 5 & 0 & 3 & -3 & 1 & 1 & 0 & 0 & -1 &
        \frac{-1 - \I \sqrt{7}}{2} & \frac{-1 + \I \sqrt{7}}{2} & -1 & 0 & 1 & 0 & 
        1 &
        -\frac{-1 - \I \sqrt{7}}{2} & -\frac{-1 + \I \sqrt{7}}{2} & 0 & 
        0 & \frac{-1 - \I \sqrt{7}}{2} & \frac{-1 + \I \sqrt{7}}{2} &
        -1 & -1
        \\
        \chi_5 &
        231 & 7 & -9 & -3 & 0 & -1 & -1 & 3 & 1 & 1 & 0 & 0 & 0 & -1
        & 1 & 0 & -1 & 0 & 0 &         0 &  \frac{-1 + \I
          \sqrt{15}}{2} & \frac{-1 - \I \sqrt{15}}{2} & 0 & 0 & 1  & 1
        \\
        \chi_6 &
        231 & 7 & -9 & -3 & 0 & -1 & -1 & 3 & 1 & 1 & 0 & 0 & 0 & -1
        & 1 & 0 & -1 & 0 & 0 &  0 & \frac{-1 - \I \sqrt{15}}{2} &
        \frac{-1 + \I \sqrt{15}}{2} & 0 & 0 & 1 & 1
        \\
        \chi_7 &
        252 & 28 & 12 & 9 & 0 & 4 & 4 & 0 & 2 & 1 & 0 & 0 & 0 & 0 &
        2 & -1 & 1 & 0 & 0 &   0 & -1 & -1 & 0 & 0 & -1 & -1
        \\
        \chi_8 &
        253 & 13 & -11 & 10 & 1 & -3 & 1 & 1 & 3 & -2  & 1  & 1 & 1
        & -1 & -1 & 0 & 0 &   1 & -1 & -1 & 0 & 0 & 1 & 1 & 0 & 0
        \\
        \chi_9 &
        483 & 35 & 3 & 6 & 0 & 3 & 3 & 3 & -2 & 2 & 0 & 0 & 0 & -1 & -2 &
        -1 & 0 & 0 & 0 & 0 &    1 & 1 & 0 & 0 & 0 & 0
        \\
        \chi_{10} &
        770 & -14 & 10 & 5 & -7 & 2 & -2 & -2 & 0 & 1 & 1 & 0 & 0 &
        0 & 0 & 0 & -1 & 1 & 0 &   0 & 0 & 0 & 0 & 0 &
        \frac{-1 + \I \sqrt{23}}{2} & \frac{-1 - \I \sqrt{23}}{2}
        \\
        \chi_{11} &
        770 & -14 & 10 & 5 & -7 & 2 & -2 & -2 & 0 & 1 & 1 & 0 & 0 &
        0 & 0 & 0 & -1 & 1 & 0 &   0 & 0 & 0 & 0 & 0 &
        \frac{-1 - \I \sqrt{23}}{2} & \frac{-1 + \I \sqrt{23}}{2}
        \\
        \chi_{12} &
        990 & -18 & -10 & 0 & 3 & 6 & 2 & -2 & 0 & 0 & -1 &
        \frac{-1 + \I \sqrt{7}}{2} & \frac{-1 - \I \sqrt{7}}{2} &
        0 & 0 & 0 & 0 & 1  &
        \frac{-1 + \I \sqrt{7}}{2} & \frac{-1 - \I \sqrt{7}}{2} & 0 &
        0 &
        \frac{-1 + \I \sqrt{7}}{2} & \frac{-1 - \I \sqrt{7}}{2} &   1
        & 1
        \\
        \chi_{13} &
        990 & -18 & -10 & 0 & 3 & 6 & 2 & -2 & 0 & 0 & -1 &
        \frac{-1 - \I \sqrt{7}}{2} & \frac{-1 + \I \sqrt{7}}{2} & 0 &
        0 & 0 & 0 & 1  &
        \frac{-1 - \I \sqrt{7}}{2} & \frac{-1 + \I \sqrt{7}}{2} & 0 &
        0 &
        \frac{-1 - \I \sqrt{7}}{2} & \frac{-1 + \I \sqrt{7}}{2} &
        1 & 1
        \\
        \chi_{14} &
        1035 & 27 & 35 & 0 & 6 & 3 & -1 & 3 & 0 & 0 & 2 & -1 & -1 & 1 & 0 & 1 & 0 & 
        0 & -1 & -1 & 0 & 0 & -1 & -1 & 0 & 0
        \\
        \chi_{15} &
        1035 & -21 & -5 & 0 & -3 & 3 & 3 & -1 & 0 & 0 &   1 & -1 +
        \I \sqrt{7} & -1 - \I \sqrt{7} & -1 & 0 & 1 & 0 & -1 & 0 & 0 &
        0 &   0 &
        -\frac{-1 + \I \sqrt{7}}{2} & -\frac{-1 - \I \sqrt{7}}{2} &
        0 & 0
        \\
        \chi_{16} &
        1035 & -21 & -5 & 0 & -3 & 3 & 3 & -1 & 0 & 0 &   1 &
        -1 -   \I \sqrt{7} & -1 + \I \sqrt{7} & -1 & 0 & 1 & 0 & -1 & 0 & 0 &
        0 &   0 &
        -\frac{-1 - \I \sqrt{7}}{2} & -\frac{-1 + \I \sqrt{7}}{2} &
        0 & 0
        \\
        \chi_{17} &
        1265 & 49 & -15 & 5 & 8 & -7 & 1 & -3 & 0 & 1 & 0 & -2 & -2
        & 1 & 0 & 0 & -1 & 0 & 0 &   0 & 0 & 0 & 1 & 1 & 0 & 0
        \\
        \chi_{18} &
        1771 & -21 & 11 & 16 & 7 & 3 & -5 & -1 & 1 & 0 & -1 & 0 & 0
        & -1 & 1 & 0 & 0 & -1  &   0 & 0 & 1 & 1 & 0 & 0 & 0 & 0
        \\
        \chi_{19} &
        2024 & 8 & 24 & -1 & 8 & 8 & 0 & 0 & -1 & -1  & 0  & 1 & 1 &
        0 & -1 & 0 & -1 & 0 & 1 &   1 & -1 & -1 & 1 & 1 & 0 & 0
        \\
        \chi_{20} &
        2277 & 21 & -19 & 0 & 6 & -3 & 1 & -3 & -3 & 0 & 2 & 2 & 2 &
        -1 & 1 & 0 & 0 & 0 & 0 &   0 & 0 & 0 & -1 & -1 & 0 & 0
        \\
        \chi_{21} &
        3312 & 48 & 16 & 0 & -6 & 0 & 0 & 0 & -3 & 0 & -2 & 1 & 1 &
        0 & 1 & 1 & 0 &   0 & -1 & -1 & 0 & 0 & 1 & 1 & 0 & 0
        \\
        \chi_{22} &
        3520 & 64 & 0 & 10 & -8 & 0 & 0 & 0 & 0 & -2 & 0 & -1 & -1 &
        0 & 0 & 0 & 0 & 0 & 1 &   1 & 0 & 0 & -1 & -1 & 1 & 1
        \\
        \chi_{23} &
        5313 & 49 & 9 & -15 & 0 & 1 & -3 & -3 & 3 & 1 & 0 & 0 & 0 &
        -1 & -1 & 0 & 1 & 0 & 0 &   0 & 0 & 0 & 0 & 0 & 0 & 0
        \\
        \chi_{24} &
        5544 & -56 & 24 & 9 & 0 & -8 & 0 & 0 & -1 & 1 & 0 & 0 & 0 &
        0 & -1 & 0 & 1 & 0 & 0 &   0 & -1 & -1 & 0 & 0 & 1 & 1
        \\
        \chi_{25} &
        5796 & -28 & 36 & -9 & 0 & -4 & 4 & 0 & 1 & -1  & 0  & 0 & 0
        & 0 & 1 & -1 & -1 & 0 &    0 & 0 & 1 & 1 & 0 & 0 & 0 & 0
        \\
        \chi_{26} &
        10395 & -21 & -45 & 0 & 0 & 3 & -1 & 3 & 0 & 0 & 0 & 0 & 0 &
        1 & 0 & 0 & 0 & 0 & 0 &   0 & 0 & 0 & 0 & 0 & -1 & -1
        \\
        \bottomrule
      \end{tabular}
    }
  }
  \caption{character table of $M_{24}$.
  $\left| M_{24} \right|=244823040$}
  \label{tab:character_M24}
\end{table}

\begin{table}[tbhp]
  \newcolumntype{L}{>{$}l<{$}}
  \newcolumntype{R}{>{$}r<{$}}
  \newcolumntype{C}{>{$}c<{$}}
  \rowcolors{2}{gray!13}{}
  \centering
      \begin{tabular}[]{RRR}
        \toprule
        g
        & \text{size} &
        \text{cycle shape}
        \\
        \midrule
        \mathrm{1A} & 1 &
        1^{24} 
        \\
        \mathrm{2A} & 11385   &
        1^8 2^8
        \\
        \mathrm{2B} & 31878   &
        2^{12}
        \\
        \mathrm{3A} & 226688  &
        1^6 3^6
        \\
        \mathrm{3B} & 485760  &
        3^8
        \\
        \mathrm{4A} & 637560  &
        2^4 4^4
        \\
        \mathrm{4B} & 1912680 & 
        1^4 2^2 4^4
        \\
        \mathrm{4C} & 2550240 & 
        4^6
        \\
        \mathrm{5A} & 4080384 &
        1^4 5^4
        \\
        \mathrm{6A} & 10200960 &
        1^2 2^2 3^2 6^2
        \\
        \mathrm{6B} & 10200960 &
        6^4
        \\
        \mathrm{7A} & 5829120 &
        1^3 7^3
        \\
        \mathrm{7B} & 5829120 &
        1^3 7^3
        \\
        \mathrm{8A} & 15301440 &
        1^2 2^1 4^1 8^2
        \\
        \mathrm{10A} & 12241152 &
        2^2 10^2
        \\
        \mathrm{11A} & 22256640 &
        1^2 11^2
        \\
        \mathrm{12A} & 20401920 &
        2^1 4^1 6^1 12^1
        \\
        \mathrm{12B} & 20401920 &
        12^2
        \\
        \mathrm{14A} & 17487360 &
        1^1 2^1 7^1 14^1
        \\
        \mathrm{14B} & 17487360 &
        1^1 2^1 7^1 14^1
        \\
        \mathrm{15A} & 16321536 &
        1^1 3^1 5^1 15^1
        \\
        \mathrm{15B} & 16321536 &
        1^1 3^1 5^1 15^1
        \\
        \mathrm{21A} & 11658240 &
        3^1 21^1
        \\
        \mathrm{21B} & 11658240 &
        3^1 21^1
        \\
        \mathrm{23A} & 10644480 &
        1^1 23^1
        \\
        \mathrm{23B} & 10644480 &
        1^1 23^1
        \\
    \bottomrule
  \end{tabular}
  \caption{Cycle shapes of conjugacy classes of $M_{24}$.}
  \label{tab:permutation_M24}
\end{table}

\begin{table}[h]
  \newcolumntype{L}{>{$}l<{$}}
  \newcolumntype{R}{>{$}r<{$}}
  \newcolumntype{C}{>{$}c<{$}}
  \rowcolors{2}{gray!13}{}
  \centering
  \rotatebox[]{90}{
    \resizebox{.93\textheight}{!}{
      \begin{tabular}[]{R|*{21}{R}}
        \toprule
        n &
        \chi_1 &
        \chi_2 &
        \chi_3 =   \chi_4 &
        \chi_5 =   \chi_6 &
        \chi_7 &
        \chi_8 &
        \chi_9 &
        \chi_{10} =   \chi_{11} &
        \chi_{12} =   \chi_{13} &
        \chi_{14} &
        \chi_{15} =   \chi_{16} &
        \chi_{17} &
        \chi_{18} &
        \chi_{19} &
        \chi_{20} &
        \chi_{21} &
        \chi_{22} &
        \chi_{23} &
        \chi_{24} &
        \chi_{25} &
        \chi_{26} 
        \\
        \midrule \midrule
  0 & -2 & 0 & 0 & 0 & 0 & 0 & 0 & 0 & 0 & 0 & 0 & 0 & 0 & 0 & 0 & 0 & 0 & 0 & 0 & 0 & 
  0 \\  1 & 0 & 0 & 1 & 0 & 0 & 0 & 0 & 0 & 0 & 0 & 0 & 0 & 0 & 0 & 0 & 0 & 0 & 0 & 0 & 0 & 
  0 \\  2 & 0 & 0 & 0 & 1 & 0 & 0 & 0 & 0 & 0 & 0 & 0 & 0 & 0 & 0 & 0 & 0 & 0 & 0 & 0 & 0 & 
  0 \\  3 & 0 & 0 & 0 & 0 & 0 & 0 & 0 & 1 & 0 & 0 & 0 & 0 & 0 & 0 & 0 & 0 & 0 & 0 & 0 & 0 & 
  0 \\  4 & 0 & 0 & 0 & 0 & 0 & 0 & 0 & 0 & 0 & 0 & 0 & 0 & 0 & 0 & 2 & 0 & 0 & 0 & 0 & 0 & 
  0 \\  5 & 0 & 0 & 0 & 0 & 0 & 0 & 0 & 0 & 0 & 0 & 0 & 0 & 0 & 0 & 0 & 0 & 0 & 0 & 0 & 2 & 
  0 \\  6 & 0 & 0 & 0 & 0 & 0 & 0 & 0 & 0 & 0 & 0 & 0 & 0 & 0 & 0 & 0 & 0 & 2 & 0 & 0 & 0 & 
  2 \\  7 & 0 & 0 & 0 & 0 & 0 & 0 & 0 & 0 & 0 & 0 & 0 & 0 & 2 & 2 & 0 & 0 & 0 & 2 & 2 & 2 & 
  2 \\  8 & 0 & 0 & 0 & 0 & 0 & 0 & 0 & 0 & 1 & 0 & 1 & 2 & 0 & 0 & 2 & 2 & 2 & 4 & 2 & 2 & 
  6 \\  9 & 0 & 0 & 0 & 0 & 0 & 0 & 2 & 2 & 0 & 2 & 2 & 0 & 2 & 2 & 2 & 4 & 4 & 4 & 8 & 8 & 
  10 \\  10 & 0 & 0 & 0 & 2 & 0 & 2 & 2 & 0 & 2 & 2 & 2 & 4 & 4 & 4 & 6 & 6 & 8 & 12 & 10 &
   10 & 24 \\  11 & 0 & 0 & 0 & 0 & 0 & 0 & 0 & 4 & 4 & 6 & 4 & 2 & 8 & 10 & 8 & 14 & 12 &
   22 & 24 & 26 & 40 \\  12 & 0 & 2 & 0 & 2 & 2 & 4 & 4 & 6 & 8 & 4 & 8 & 12 & 12 & 12 & 
  18 & 26 & 30 & 40 & 38 & 40 & 80 \\  13 & 0 & 0 & 2 & 2 & 4 & 2 & 6 & 10 & 14 & 18 & 
  14 & 16 & 26 & 30 & 28 & 44 & 44 & 70 & 80 & 84 & 136 \\  14 & 0 & 0 & 0 & 8 & 4 & 6 &
   14 & 16 & 24 & 22 & 24 & 34 & 38 & 46 & 58 & 80 & 86 & 128 & 126 & 132 & 
  254 \\  15 & 0 & 0 & 2 & 8 & 12 & 8 & 18 & 38 & 40 & 46 & 44 & 46 & 78 & 86 & 88 & 
  138 & 144 & 218 & 238 & 246 & 424 \\  16 & 0 & 2 & 2 & 18 & 18 & 22 & 36 & 50 & 72 &
   68 & 72 & 100 & 122 & 140 & 170 & 232 & 252 & 378 & 382 & 400 & 742 \\  17 & 0 & 
  2 & 8 & 25 & 30 & 26 & 54 & 94 & 116 & 130 & 124 & 140 & 212 & 246 & 262 & 392 & 
  410 & 630 & 670 & 704 & 1222 \\  18 & 0 & 6 & 6 & 50 & 50 & 58 & 100 & 148 & 194 & 
  192 & 202 & 256 & 342 & 388 & 454 & 654 & 704 & 1044 & 1074 & 1120 & 
  2058 \\  19 & 0 & 4 & 18 & 68 & 80 & 72 & 150 & 252 & 318 & 346 & 332 & 394 & 582 &
   664 & 722 & 1062 & 1116 & 1702 & 1800 & 1880 & 3320 \\  20 & 0 & 14 & 20 & 126 &
   128 & 138 & 254 & 390 & 516 & 520 & 536 & 676 & 904 & 1036 & 1196 & 1716 & 
  1836 & 2764 & 2846 & 2980 & 5408 \\  21 & 2 & 20 & 40 & 182 & 214 & 200 & 396 & 
  652 & 814 & 872 & 860 & 1020 & 1476 & 1684 & 1862 & 2742 & 2902 & 4384 & 4622 & 
  4828 & 8572 \\  22 & 2 & 32 & 55 & 314 & 328 & 346 & 640 & 988 & 1298 & 1336 & 
  1348 & 1686 & 2302 & 2630 & 3000 & 4324 & 4616 & 6950 & 7204 & 7532 & 
  13620 \\  23 & 2 & 40 & 98 & 460 & 512 & 496 & 972 & 1590 & 2020 & 2144 & 2118 & 
  2546 & 3638 & 4162 & 4624 & 6768 & 7166 & 10856 & 11376 & 11898 & 
  21204 \\  24 & 0 & 80 & 132 & 744 & 798 & 824 & 1544 & 2426 & 3140 & 3236 & 
  3278 & 4050 & 5584 & 6376 & 7248 & 10500 & 11192 & 16834 & 17504 & 18294 & 
  32976 \\  25 & 8 & 108 & 234 & 1106 & 1232 & 1208 & 2336 & 3764 & 4814 & 5084 & 
  5038 & 6108 & 8654 & 9892 & 11042 & 16112 & 17084 & 25840 & 27056 & 28288 & 
  50524 \\  26 & 6 & 174 & 322 & 1742 & 1860 & 1904 & 3602 & 5677 & 7348 & 7626 & 
  7670 & 9444 & 13090 & 14968 & 16940 & 24566 & 26148 & 39428 & 41022 & 42894 & 
  77176 \\  27 & 12 & 252 & 514 & 2560 & 2836 & 2802 & 5394 & 8688 & 11092 & 
  11666 & 11618 & 14100 & 19914 & 22744 & 25462 & 37148 & 39436 & 59564 & 
  62294 & 65114 & 116494 \\  28 & 16 & 398 & 742 & 3922 & 4238 & 4310 & 8160 & 
  12912 & 16686 & 17356 & 17418 & 21414 & 29772 & 34026 & 38434 & 55764 & 
  59330 & 89490 & 93218 & 97456 & 175146 \\  29 & 26 & 560 & 1154 & 5758 & 6328 &
   6286 & 12090 & 19380 & 24840 & 26078 & 25994 & 31636 & 44512 & 50892 & 
  57068 & 83146 & 88280 & 133356 & 139342 & 145690 & 260828 \\  30 & 34 & 876 & 
  1642 & 8642 & 9368 & 9486 & 18008 & 28580 & 36824 & 38368 & 38480 & 47172 & 
  65776 & 75158 & 84776 & 123176 & 131020 & 197596 & 205970 & 215318 & 
  386724 
        \\
        \bottomrule
      \end{tabular}
    }}
  \caption{multiplicities of the decomposition of $A(n)$ into irreducible representations of  $M_{24}$ in Mathieu moonshine}
  \label{tab:multiplicity_M24}
\end{table}


\begin{table}[h]
  \newcolumntype{L}{>{$}l<{$}}
  \newcolumntype{R}{>{$}r<{$}}
  \newcolumntype{C}{>{$}c<{$}}
  \rowcolors{2}{gray!13}{}
  \centering
  \rotatebox[]{90}{
    \resizebox{.77\textheight}{!}{
      \begin{tabular}[]{R|*{21}{R}}
        \toprule
        n &
        \mathrm{1A}&        
        \mathrm{2A}&
        \mathrm{2B}&  
        \mathrm{3A} & \mathrm{3B}&
        \mathrm{4A} &
        \mathrm{4B}& \mathrm{4C}&
        \mathrm{5A}&                
        \mathrm{6A}& \mathrm{6B}&
        \mathrm{7AB}&
        \mathrm{8A}&
        \mathrm{10A} &
        \mathrm{11A}&
        \mathrm{12A} & \mathrm{12B} &
        \mathrm{14AB} & 
        \mathrm{15AB} &
        \mathrm{21AB} &
        \mathrm{23AB}
        \\
    \midrule \midrule
0 & -2 & -2 & -2 & -2 & -2 & -2 & -2 & -2 & -2 & -2 & -2 & -2 & -2 & -2 & -2 & -2 &  
-2 & -2 & -2 & -2 & -2\\ 1 & 90 & -6 & 10 & 0 & 6 & -6 & 2 & 2 & 0 & 0 & -2 & -1 & -2 &
   0 & 2 & 0 & 2 & 1 & 0 & -1 & -2\\ 2 & 462 & 14 & -18 & -6 & 0 & -2 & -2 & 6 & 2 & 
  2 & 0 & 0 & -2 & 2 & 0 & -2 & 0 & 0 & -1 & 0 & 2\\ 3 & 1540 & -28 & 20 & 10 & -14 & 
  4 & -4 & -4 & 0 & 2 & 2 & 0 & 0 & 0 & 0 & -2 & 2 & 0 & 0 & 0 & -1\\ 4 & 4554 & 
  42 & -38 & 0 & 12 & -6 & 2 & -6 & -6 & 0 & 4 & 4 & -2 & 2 & 0 & 0 & 0 & 0 & 0 & -2 & 
  0\\ 5 & 11592 & -56 & 72 & -18 & 0 & -8 & 8 & 0 & 2 & -2 & 0 & 0 & 0 & 2 & -2 & -2 &
   0 & 0 & 2 & 0 & 0\\ 6 & 27830 & 86 & -90 & 20 & -16 & 6 & -2 & 6 & 0 & -4 & 
  0 & -2 & 2 & 0 & 0 & 0 & 0 & 2 & 0 & -2 & 0\\ 7 & 61686 & -138 & 118 & 0 & 30 & 
  6 & -10 & -2 & 6 & 0 & -2 & 2 & -2 & -2 & -2 & 0 & -2 & 2 & 0 & 2 & 0\\ 8 & 131100 &
   188 & -180 & -30 & 0 & -4 & 4 & -12 & 0 & 2 & 0 & -3 & 0 & 0 & 2 & 2 & 0 & -1 & 0 & 
  0 & 0\\ 9 & 265650 & -238 & 258 & 42 & -42 & -14 & 10 & 10 & -10 & 2 & 6 & 
  0 & -2 & -2 & 0 & -2 & -2 & 0 & 2 & 0 & 0\\ 10 & 521136 & 336 & -352 & 0 & 42 & 
  0 & -8 & 16 & 6 & 0 & 2 & 0 & -4 & -2 & 0 & 0 & -2 & 0 & 0 & 0 & 2\\ 11 & 
  988770 & -478 & 450 & -60 & 0 & 18 & -14 & -6 & 0 & -4 & 0 & 6 & 2 & 0 & 2 & 0 & 
  0 & -2 & 0 & 0 & 0\\ 12 & 1830248 & 616 & -600 & 62 & -70 & -8 & 8 & -16 & 
  8 & -2 & -6 & 0 & 0 & 0 & 2 & -2 & 2 & 0 & 2 & 0 & 0\\ 13 & 3303630 & -786 & 830 & 
  0 & 84 & -18 & 22 & 6 & 0 & 0 & -4 & -6 & 2 & 0 & 0 & 0 & 0 & -2 & 0 & 0 & 2\\ 14 & 
  5844762 & 1050 & -1062 & -90 & 0 & 10 & -6 & 18 & -18 & 6 & 0 & 0 & 2 & -2 & 
  0 & -2 & 0 & 0 & 0 & 0 & 2\\ 15 & 10139734 & -1386 & 1334 & 118 & -110 & 
  22 & -26 & -10 & 4 & 6 & 2 & -4 & -2 & 4 & 0 & -2 & 2 & 0 & -2 & 2 & 0\\ 16 & 
  17301060 & 1764 & -1740 & 0 & 126 & -12 & 12 & -28 & 0 & 0 & 6 & 0 & 0 & 0 & -4 & 
  0 & 2 & 0 & 0 & 0 & 0\\ 17 & 29051484 & -2212 & 2268 & -156 & 0 & -36 & 28 & 12 &
   14 & -4 & 0 & 0 & -4 & -2 & 0 & 0 & 0 & 0 & -1 & 0 & 0\\ 18 & 48106430 & 
  2814 & -2850 & 170 & -166 & 14 & -18 & 38 & 0 & -6 & -6 & 8 & -2 & 0 & -2 & 2 & 2 & 
  0 & 0 & 2 & -2\\ 19 & 78599556 & -3612 & 3540 & 0 & 210 & 36 & -36 & -20 & -24 &
   0 & -6 & 0 & 0 & 0 & 2 & 0 & -2 & 0 & 0 & 0 & 0\\ 20 & 126894174 & 
  4510 & -4482 & -228 & 0 & -18 & 14 & -42 & 14 & 4 & 0 & -6 & -2 & -2 & 0 & 0 & 0 & 
  2 & 2 & 0 & 0\\ 21 & 202537080 & -5544 & 5640 & 270 & -282 & -40 & 48 & 16 & 0 &
   6 & 6 & 4 & 4 & 0 & -2 & 2 & -2 & 0 & 0 & -2 & 0\\ 22 & 319927608 & 
  6936 & -6968 & 0 & 300 & 24 & -16 & 48 & 18 & 0 & 4 & -7 & 4 & 2 & 0 & 0 & 0 & -1 & 
  0 & -1 & 0\\ 23 & 500376870 & -8666 & 8550 & -360 & 0 & 54 & -58 & -18 & 
  0 & -8 & 0 & 0 & -2 & 0 & 4 & 0 & 0 & 0 & 0 & 0 & 2\\ 24 & 775492564 & 
  10612 & -10556 & 400 & -392 & -28 & 28 & -60 & -36 & -8 & -8 & 0 & 0 & 4 & 0 & -4 &
   0 & 0 & 0 & 0 & 0\\ 25 & 1191453912 & -12936 & 13064 & 0 & 462 & -72 & 64 & 
  32 & 12 & 0 & -10 & 12 & -4 & 4 & 0 & 0 & 2 & 0 & 0 & 0 & 0\\ 26 & 1815754710 & 
  15862 & -15930 & -510 & 0 & 22 & -34 & 78 & 0 & 10 & 0 & 0 & -6 & 0 & 0 & -2 & 0 & 
  0 & 0 & 0 & -1\\ 27 & 2745870180 & -19420 & 19268 & 600 & -600 & 
  84 & -76 & -36 & 30 & 8 & 8 & -10 & 4 & -2 & -2 & 0 & 0 & -2 & 0 & 2 & 0\\ 28 & 
  4122417420 & 23532 & -23460 & 0 & 660 & -36 & 36 & -84 & 0 & 0 & 12 & 2 & 0 & 0 & 
  0 & 0 & 0 & -2 & 0 & 2 & 0\\ 29 & 6146311620 & -28348 & 28548 & -762 & 0 & -92 &
   100 & 36 & -50 & -10 & 0 & -6 & 4 & -2 & -2 & -2 & 0 & 2 & -2 & 0 & 0\\ 30 & 
  9104078592 & 34272 & -34352 & 828 & -840 & 48 & -40 & 96 & 22 & -12 & -8 & 0 & 
  4 & -2 & 4 & 0 & 0 & 0 & -2 & 0 & 0
    \\
    \bottomrule
      \end{tabular}
    }
}
  \caption{Expansion coefficients of $A_g(n)$ in Mathieu moonshine.}
  \label{tab:coeff_M24}
\end{table}



\begin{table}[tbhp]
  \newcolumntype{L}{>{$}l<{$}}
  \newcolumntype{R}{>{$}r<{$}}
  \newcolumntype{C}{>{$}c<{$}}
  \rowcolors{2}{gray!13}{}
  \centering
  \begin{tabular}[]{RRR}
    \toprule
    g
    & \text{size} &
    \text{cycle shape}
    \\
    \midrule
    \mathrm{1A} & 1 &
    1^{12}
    \\
    \mathrm{2A} & 396 &
    2^6
    \\
    \mathrm{2B} & 495 &
    1^4 2^4
    \\
    \mathrm{3A} & 1760 &
    1^3 3^3
    \\
    \mathrm{3B} & 2640 &
    3^4
    \\
    \mathrm{4A} & 2970 &
    2^2 4^2
    \\
    \mathrm{4B} & 2970 & 
    1^4 4^2
    \\
    \mathrm{5A} & 9504 &
    1^2 5^2
    \\
    \mathrm{6A} & 7920 &
    6^2
    \\
    \mathrm{6B} & 15840 &
    1^1 2^1 3^1 6^1
    \\
    \mathrm{8A} & 11880 &
    4^1 8^1
    \\
    \mathrm{8B} & 11880 &
    1^2 2^1 8^1
    \\
    \mathrm{10A} & 9504 &
    2^1 10^1
    \\
    \mathrm{11A} & 8640 &
    1^1 11^1
    \\
    \mathrm{11B} & 8640 &
    1^1 11^1
    \\
    \bottomrule
  \end{tabular}
  \caption{Cycle shapes of conjugacy classes of
    $M_{12}$.}
  \label{tab:permutation_M12}
\end{table}
\begin{table}[h]
  \newcolumntype{L}{>{$}l<{$}}
  \newcolumntype{R}{>{$}r<{$}}
  \newcolumntype{C}{>{$}c<{$}}
  \rowcolors{2}{gray!13}{}
  \centering
      \begin{tabular}[]{R|*{15}{R}}
        \toprule
        &
        \mathrm{1A}&        
        \mathrm{2A}&
        \mathrm{2B}&  
        \mathrm{3A} & \mathrm{3B}&
        \mathrm{4A} &
        \mathrm{4B}&                 \mathrm{5A}&                
        \mathrm{6A}& \mathrm{6B}&
        \mathrm{8A}& 
        \mathrm{8B}  &\mathrm{10A}
        &\mathrm{11A}&\mathrm{11B}
        \\
        \midrule \midrule
        \chi_1 &
        1 &1 &1 &1   &1 &1  &1 &1 &1 &1
        &1 &1&1&1 & 1
        \\
        \chi_2 &
        11&-1&3&2&-1&-1&3&1 &-1&0&-1&1
        &-1&0&0
        \\
        \chi_3 &
        11&-1&3&2&-1&3&-1&1 &-1&0&1&-1
        &-1&0&0
        \\
        \chi_4 &
        16&4&0&-2&1&0&0&1 &1 &0&0&0
        &-1&\frac{-1+\I\sqrt{11}}{2}&\frac{-1-\I\sqrt{11}}{2}
        \\
        \chi_5 &
        16&4&0&-2&1&0&0&1 &1&0&0&0
        &-1&\frac{-1-\I\sqrt{11}}{2}&\frac{-1+\I\sqrt{11}}{2}
        \\
        \chi_6 &
        45&5&-3&0&3&1&1&0 &-1&0&-1&-1
        &0&1&1
        \\
        \chi_7 &
        54&6&6&0&0&2&2&-1 &0&0&0&0
        &1&-1&-1
        \\
        \chi_8 &
        55&-5&7&1&1&-1&-1 &0& 1&1&-1&-1
        &0&0&0
        \\
        \chi_9 &
        55&-5&-1&1&1&3&-1 &0& 1&-1&-1&1
        &0&0&0
        \\
        \chi_{10} &
        55&-5&-1&1&1&-1&3 &0 &1&-1&1&-1
        &0&0&0
        \\
        \chi_{11} &
        66&6&2&3&0&-2&-2 &1& 0&-1&0&0
        &1&0&0
        \\
        \chi_{12} &
        99&-1&3&0&3&-1&-1&-1&-1&0&1&1
        &-1&0&0
        \\
        \chi_{13} &
        120&0&-8&3&0&0&0 &0 &0&1&0&0
        &0&-1&-1
        \\
        \chi_{14} &
        144&4&0&0&-3&0&0&-1 &1&0&0&0
        &-1&1&1
        \\
        \chi_{15} &
        176&-4&0&-4&-1&0&0&1&-1&0&0&0
        &1&0&0
        \\
        \bottomrule
      \end{tabular}
  \caption{character table of $M_{12}$.
  $\left| M_{12} \right|=95040$}
  \label{tab:character_M12}
\end{table}

\begin{table}[h]
  \newcolumntype{L}{>{$}l<{$}}
  \newcolumntype{R}{>{$}r<{$}}
  \newcolumntype{C}{>{$}c<{$}}
  \rowcolors{2}{gray!13}{}
  \centering
      \begin{tabular}[]{CC|*{15}{C}}
        \toprule
        \multicolumn{2}{C|}{
          M_{24} \backslash M_{12}
        } &
        \chi_1 &
        \chi_2 &
        \chi_3 &
        \chi_4 &
        \chi_5 &
        \chi_6 &
        \chi_7 &
        \chi_8 &
        \chi_9 &
        \chi_{10} &
        \chi_{11} &
        \chi_{12} &
        \chi_{13} &
        \chi_{14} &
        \chi_{15} 
        \\
        \multicolumn{2}{C|}{ }&
        1 & 11 & 11 & 16 & 16 & 45 & 54 & 55 & 55 & 55 & 66 & 
        99 & 120 & 144 & 176
        \\
        \midrule \midrule
        \chi_1 & 1 &
        1 & & & & & & & & & & & & & &
        \\
        \chi_2 & 23 &
        1 & 1 & 1 & & & & & & & & & & & &
        \\
        \chi_3 & 45 &
        & & & & & 1 & & & & & & & & &
        \\
        \chi_4 & 45 &
        & & & & & 1 & & & & & & & & &
        \\
        \chi_5 & 231 &
        & & & & &  & &1 & & & & & & & 1
        \\
        \chi_6 & 231 &
        & & & & &  & & 1& & & & & & & 1
        \\
        \chi_7 & 252 &
        1 & 1 & 1 & & & &2 &1 & & &1 & & & &
        \\
        \chi_8 & 253 &
        & 1& 1& & & & &1 &1 &1 &1 & & & &
        \\
        \chi_9 & 483 &
        &1 &1 & & & & 2& 2& & & &1 & &1 &
        \\
        \chi_{10} & 770 &
        & & & & & & & & & &1 & & 2& 2&1
        \\
        \chi_{11} & 770 &
        & & & & & & & & & &1 & & 2& 2&1
        \\
        \chi_{12} & 990 &
        & & & & &1 & & & 1& 1& & 1& 2& 1&2
        \\
        \chi_{13} & 990 &
        & & & & &1 & & & 1& 1& & 1& 2& 1&2
        \\
        \chi_{14} & 1035 &
        1& & &1 &1 &1 &2 &1 & & &2 &2 & &2 &1
        \\
        \chi_{15} & 1035 &
        & & & & & 1& & & 1& 1& & & 2& 2&2
        \\
        \chi_{16} & 1035 &
        & & & & & 1& & & 1& 1& & & 2& 2&2
        \\
        \chi_{17} &1265 &
        1&1 &1 & & & & 2& 3& 1& 1& 1& 3& & 1&2
        \\
        \chi_{18} & 1771&
        & & & & & 2& &1 &1 &1 &3 &2 &4 &2 &2
        \\
        \chi_{19} & 2024 &
        & & &1 &1 &2 &2 &1 &1 &1 &2 &3 &2 &3 &3
        \\
        \chi_{20} & 2277 &
        & & & & &1 &2 &3 &2 &2 &1 &3 &2 &3 &4
        \\
        \chi_{21} & 3312 &
        & 1& 1& & & 1& 4& 3& 1& 1& 3&4 &2 &6 &6
        \\
        \chi_{22} & 3520 &
        & 2& 2& & & &4 &4 &2 &2 &4 &4 &2 &6 &6
        \\
        \chi_{23} & 5313&
        &1 &1 &2 &2 &2 &4 &5 &2 &2 &4 &6 &4 &8 &11
        \\
        \chi_{24} & 5544 &
        & & &1 &1 &4 &2 &1 &3 &3 &4 &5 &10 &9 &9
        \\
        \chi_{25} & 5796&
        & & &2 &2 &4 &4 &1 &3 &3 &4 &5 &8 &9 &11
        \\
        \chi_{26} & 10395&
        & 1& 1& 1& 1& 4& 4& 6& 7& 7& 6& 11&14 &15 &20
        \\
        \bottomrule
      \end{tabular}
    \caption{Branching of $M_{24}$ representations into those of
      $M_{12}$. Only 
      non-zero multiplicities are written.
    }
    \label{tab:branching}
  \end{table}
      
\begin{table}[h]
  \newcolumntype{L}{>{$}l<{$}}
  \newcolumntype{R}{>{$}r<{$}}
  \newcolumntype{C}{>{$}c<{$}}
  \rowcolors{2}{gray!13}{}
  \centering
    \resizebox{.97\textwidth}{!}{
      \begin{tabular}[]{R|*{12}{R}}
        \toprule
        n &
        \chi_1 &
        \chi_2 =     \chi_3 &
        \chi_4 =     \chi_5 &
        \chi_6 &
        \chi_7 &
        \chi_8 &
        \chi_9 =   \chi_{10} &
        \chi_{11} &
        \chi_{12} &
        \chi_{13} &
        \chi_{14} &
        \chi_{15} 
        \\
        \midrule \midrule
  0 & -1 & 0 & 0 & 0 & 0 & 0 & 0 & 0 & 0 & 0 & 0 & 0 \\  1 & 0 & 0 & 0 & 1 & 0 & 0 & 0 & 0 &
   0 & 0 & 0 & 0 \\  2 & 0 & 0 & 0 & 0 & 0 & 1 & 0 & 0 & 0 & 0 & 0 & 1 \\  3 & 0 & 0 & 0 & 
  0 & 0 & 0 & 0 & 1 & 0 & 2 & 2 & 1 \\  4 & 0 & 0 & 0 & 1 & 2 & 3 & 2 & 1 & 3 & 2 & 3 & 
  4 \\  5 & 0 & 0 & 2 & 4 & 4 & 1 & 3 & 4 & 5 & 8 & 9 & 11 \\  6 & 0 & 3 & 1 & 4 & 8 & 10 &
   9 & 10 & 15 & 16 & 21 & 26 \\  7 & 0 & 2 & 7 & 18 & 16 & 15 & 17 & 23 & 32 & 42 & 
  46 & 56 \\  8 & 1 & 9 & 10 & 28 & 38 & 43 & 39 & 43 & 70 & 78 & 98 & 124 \\  9 & 1 & 
  14 & 23 & 66 & 76 & 70 & 75 & 94 & 134 & 174 & 206 & 242 \\  10 & 3 & 33 & 42 & 
  119 & 148 & 162 & 154 & 179 & 276 & 322 & 390 & 485 \\  11 & 4 & 51 & 88 & 242 & 
  278 & 272 & 282 & 346 & 511 & 632 & 753 & 914 \\  12 & 10 & 115 & 147 & 420 & 
  522 & 546 & 534 & 633 & 956 & 1144 & 1384 & 1699 \\  13 & 19 & 183 & 286 & 801 & 
  938 & 933 & 951 & 1152 & 1716 & 2102 & 2506 & 3051 \\  14 & 30 & 346 & 484 & 
  1364 & 1664 & 1721 & 1698 & 2018 & 3056 & 3666 & 4420 & 5423 \\  15 & 52 & 576 &
   861 & 2420 & 2874 & 2896 & 2922 & 3535 & 5263 & 6434 & 7697 & 9375 \\  16 & 
  94 & 1017 & 1444 & 4069 & 4922 & 5058 & 5022 & 5994 & 9033 & 10886 & 13087 & 
  16032 \\  17 & 151 & 1658 & 2468 & 6920 & 8248 & 8340 & 8388 & 10099 & 15107 & 
  18382 & 22027 & 26887 \\  18 & 252 & 2817 & 4020 & 11330 & 13674 & 14000 & 
  13941 & 16689 & 25077 & 30316 & 36427 & 44563 \\  19 & 412 & 4508 & 6647 & 
  18681 & 22316 & 22644 & 22717 & 27318 & 40913 & 49696 & 59567 & 72744 \\  20 &
   669 & 7385 & 10649 & 29960 & 36064 & 36844 & 36750 & 44021 & 66134 & 80010 & 
  96094 & 117541 \\  21 & 1064 & 11676 & 17087 & 48040 & 57526 & 58442 & 58560 &
   70371 & 105420 & 127988 & 153496 & 187481 \\  22 & 1692 & 18579 & 26877 & 
  75625 & 90908 & 92775 & 92630 & 111037 & 166710 & 201830 & 242298 & 
  296284 \\  23 & 2622 & 28863 & 42197 & 118616 & 142120 & 144536 & 144714 & 
  173798 & 260529 & 316064 & 379145 & 463254 \\  24 & 4082 & 44995 & 65174 & 
  183384 & 220348 & 224690 & 224472 & 269200 & 403992 & 489368 & 587424 & 
  718126 \\  25 & 6270 & 68818 & 100406 & 282327 & 338446 & 344382 & 344655 & 
  413792 & 620437 & 752450 & 902705 & 1103084 \\  26 & 9555 & 105225 & 152718 &
   429576 & 515886 & 525845 & 525510 & 630341 & 945863 & 1145966 & 1375439 & 
  1681406 \\  27 & 14433 & 158731 & 231277 & 650388 & 780008 & 793968 & 
  794367 & 953589 & 1429925 & 1733926 & 2080389 & 2542299 \\  28 & 21711 & 
  238790 & 346819 & 975551 & 1171218 & 1193511 & 1193023 & 1431222 & 2147351 &
   2602046 & 3122821 & 3817239 \\  29 & 32314 & 355395 & 517616 & 1455614 & 
  1746034 & 1777621 & 1778220 & 2134316 & 3200923 & 3880816 & 4656537 & 
  5690817 \\  30 & 47909 & 527223 & 766024 & 2154660 & 2586488 & 2635260 & 
  2634546 & 3160915 & 4742013 & 5746832 & 6896777 & 8429971
        \\
        \bottomrule
      \end{tabular}
    }
  \caption{multiplicities of irreducible representations of $M_{12}$ in Enriques moonshine}
  \label{tab:multiplicity_M12}
\end{table}

  
\clearpage


\begin{thebibliography}{10}
\providecommand{\url}[1]{\texttt{#1}}
\providecommand{\urlprefix}{URL }
\providecommand{\eprint}[2][]{\url{#2}}

\bibitem{MCheng10a}
M.~C.~N. Cheng, \emph{{$K3$} surfaces, {$\mathcal{N}=4$} dyons, and the
  {Mathieu} group {$M_{24}$}}, Commun. Number Theory Phys. \textbf{4}, 623--657
  (2010), \href{http://jp.arxiv.org/abs/1005.5415}{\texttt{arXiv:1005.5415
  [hep-th]}}.

\bibitem{ChenDuncHarv12a}
M.~C.~N. Cheng, J.~F.~R. Duncan, and J.~A. Harvey, \emph{Umbral moonshine},
  preprint  (2012),
  \href{http://jp.arxiv.org/abs/1204.2779}{\texttt{arXiv:1204.2779 [math.RT]}}.

\bibitem{Conway98a}
J.~H. Conway, \emph{Three lectures on exceptional groups}, in J.~H. Conway and
  N.~J.~A. Sloane, eds., \emph{Sphere Packings, Lattices and Groups}, vol. 290
  of \emph{Grund. math. Wiss.}, chap.~10, pp. 267--298, Springer, Berlin, 1998.

\bibitem{EguchiHikami09b}
T.~Eguchi and K.~Hikami, \emph{{$\mathcal{N}=4$} superconformal algebra and the
  entropy of hyper{K\"ahler} manifolds},
  \href{http://dx.doi.org/10.1007/JHEP02(2010)019}{J. High Energy Phys.}
  \textbf{2010:02}, 019 (2010), 28 pages,
  \href{http://jp.arxiv.org/abs/0909.0410}{\texttt{arXiv:0909.0410 [hep-th]}}.

\bibitem{EguchiHikami10b}
---{}---{}---, \emph{Note on twisted elliptic genus of {$K3$} surface},
  \href{http://dx.doi.org/10.1016/j.physletb.2010.10.017}{Phys. Lett. B}
  \textbf{694}, 446--455 (2011),
  \href{http://jp.arxiv.org/abs/1008.4924}{\texttt{arXiv:1008.4924 [hep-th]}}.

\bibitem{EguchiHikami12a}
---{}---{}---, \emph{{$\mathcal{N}=2$} moonshine},
  \href{http://dx.doi.org/10.1016/j.physletb.2012.09.037}{Phys. Lett. B}
  \textbf{717}, 266--273 (2012),
  \href{http://jp.arxiv.org/abs/1209.0610}{\texttt{arXiv:1209.0610 [hep-th]}}.

\bibitem{EguchiHikami13}
---{}---{}---,
 in preparation.


\bibitem{EgucOoguTach10a}
T.~Eguchi, H.~Ooguri, and Y.~Tachikawa, \emph{Notes on the {$K3$} surface and
  the {Mathieu} group {$M_{24}$}},
  \href{http://dx.doi.org/10.1080/10586458.2011.544585}{Exp. Math.}
  \textbf{20}, 91--96 (2011),
  \href{http://jp.arxiv.org/abs/1004.0956}{\texttt{arXiv:1004.0956 [hep-th]}}.

\bibitem{GabeHoheVolp10a}
M.~R. Gaberdiel, S.~Hohenegger, and R.~Volpato,
\emph{{Mathieu} twining characters for {$K3$}},
\href{http://dx.doi.org/10.1007/JHEP09(2010)058}{J. High Energy Phys.}
\textbf{2010:09}, 058 (2010), 20 pages,
\href{http://jp.arxiv.org/abs/1006.0221}{\texttt{arXiv:1006.0221 [hep-th]}}.

\bibitem{GabeHoheVolp10b}
  ---{}---{}---,
  \emph{{Mathieu} moonshine in
    the elliptic genus of {$K3$}},
  \href{http://dx.doi.org/10.1007/JHEP10(2010)062}{J. High Energy Phys.}
  \textbf{2010:10}, 062 (2010), 24 pages,
  \href{http://jp.arxiv.org/abs/1008.3778}{\texttt{arXiv:1008.3778 [math.AG]}}.


\bibitem{Gannon12a}
T.~Gannon, \emph{Much ado about {Mathieu}}, preprint  (2012),
  \href{http://jp.arxiv.org/abs/1211.5531}{\texttt{arXiv:1211.5531 [math.RT]}}.

\bibitem{Mukai12a}
S.~Mukai, \emph{Lecture notes on {$K3$} and {Enriques} surfaces}, in
  P.~Pragacz, ed., \emph{Contributions to Algebraic Geometry}, EMS Series of
  Congress Reports, pp. 389--405, European Math. Soc., Z{\"u}rich, 2012,
  impanga lecture notes.


\bibitem{Govind10b}
S.~Govindarajan, \emph{Brewing moonshine for  {Mathieu}}, preprint  (2010),
  \href{http://jp.arxiv.org/abs/1211.5531}{\texttt{arXiv:1012.5732 [hep-th]}}.

\end{thebibliography}

\end{document}